\documentclass[prl,twocolumn,superscriptaddress,showpacs,preprintnumbers,amsmath,amssymb]{revtex4}

\usepackage{graphicx}
\usepackage{dcolumn}
\usepackage{bm}
\usepackage{ifpdf}

\begin{document}

\bibliographystyle{apsrev} 


\title{Localization of spin mixing dynamics in a spin-1 Bose-Einstein condensate}

\author{Wenxian Zhang}
\affiliation{The Key Laboratory for Advanced Materials and Devices,
 Department of Optical Science and Engineering, Fudan University, Shanghai 200433, China}

\author{Bo Sun}
\affiliation{Department of Physics, Auburn University, Auburn,
Alabama 36849, USA}

\author{M. S. Chapman}
\affiliation{School of Physics, Georgia Institute of Technology,
Atlanta, Georgia 30332-0430, USA}

\author{L. You}
\affiliation{School of Physics, Georgia Institute of Technology,
Atlanta, Georgia 30332-0430, USA}

\date{\today}

\begin{abstract}
We propose to localize spin mixing dynamics in a spin-1
Bose-Einstein condensate by a temporal modulation of spin exchange
interaction, which is tunable with optical Feshbach resonance.
Adopting techniques from coherent control, we
demonstrate the localization/freezing of spin mixing dynamics,
and the suppression of the intrinsic dynamic instability
and spontaneous spin domain formation in
a ferromagnetically interacting condensate of $^{87}$Rb atoms.
This work points to a promising scheme for investigating
the weak magnetic spin dipole interaction, which is usually
masked by the more dominant spin exchange interaction.
\end{abstract}


\maketitle

Dynamic localization is ubiquitous in nonlinear systems,
both for classical dynamics as in an inverted pendulum
with a rapidly modulating pivot~\cite{LandauBook60}
or an ion in a Paul trap~\cite{Horvath97},
and for quantum dynamics like a one- or
two-dimensional soliton in a Bose-Einstein
condensate (BEC) when the attractive interaction strength
is rapidly modulated \cite{Saito03, Abdullaev00, Abdullaev03, Montesinos04}.
It is often used
to stabilize a dynamically unstable system.

Spin mixing dynamics of a spin-1 atomic condensate
are dynamically unstable~\cite{Zhang05b} when the spin exchange interaction
is ferromagnetic, {\it i.e.}, favoring a ground state with all atomic spins
aligned. When confined spatially, the unstable dynamics is known to cause
formation of spin domain structures~\cite{Chang05, Sadler06}.
For many applications of spinor condensates,
from quantum simulation to precision measurement~\cite{Vengalattore07},
it is desirable that spin domain formation is suppressed.
In addition, atomic spin dipolar interactions, although weak
when compared to typical spin exchange interactions,
induce intricate spin textures that are
difficult to probe when masked by spin domain structures.
Thus the suppressing/freezing of the undesirable dynamics
from spin exchange interaction is also important for
investigating the effect of dipolar interaction \cite{Yi06,Vengalattore08}.

Compared to conventional magnets in solid states, a spin-1 BEC has
one unsurpassed advantage: its spin exchange interaction between
individual atoms can be precisely tuned through optical (as well as magnetic)
Feshbach resonances~\cite{Fatemi00, Theis04, Theis05, Cheng08, Hamley09, Jing09, Jack05}.
By adjusting the two $s$-wave scattering lengths $a_0$ and $a_2$ of two
colliding spin-1 atoms via optical means, the spin exchange interaction
strength, characterized by $c_2 =4\pi \hbar^2 (a_2-a_0)/3M$ with $M$ the mass of the atom,
is tunable. Analogous to an inverted rigid pendulum with a rapidly oscillating pivot,
a fast temporal modulation of the spin exchange interaction can localize
the spin mixing dynamics, equivalent to a suppressing/nulling of the spin exchange interaction.

This study is devoted to a theoretical investigation of spin dynamics
in a spin-1 BEC under the temporal modulation of the spin exchange interaction.
As an application, we illustrate the suppression of the dynamic instability
and the resulting prevention of spin domain formation in a condensate
with ferromagnetic interaction. The proposed scheme to control the spin exchange
interaction will potentially provide a substantial improvement
to the accuracy of several envisaged magnetometer setups and
to enable cleaner detections of dipolar effects.

For both spin-1 atoms $^{87}$Rb and $^{23}$Na, popular experimental choices,
their spin-independent interaction strength, characterized by $c_0 = 4\pi \hbar^2 (2a_2+a_0)/3M$,
is two to three orders of magnitude larger than $|c_2|$~\cite{Ho98, Law98, Ohmi98}.
This ensures the validity of single spatial mode approximation (SMA) \cite{Yi02, Zhang03, Black07, Liu09a, Liu09b}
when the number of atoms is small and the magnetic field is low.
The spin degrees of freedom and the spatial degrees
of freedom become separated within the SMA. This allows one
focus on the most interesting spin dynamics free from density dependent interactions.

Within the mean field framework, the spin dynamics of a spin-1 condensate
under the SMA is described by~\cite{Zhang05a}
\begin{eqnarray}
\label{eq:sma}
\dot \rho_0 &=& {2c\over \hbar}\rho_0 \sqrt{(1-\rho_0)^2-m^2} \sin \theta, \\
\dot \theta &=& {2c\over \hbar}\left[(1-2\rho_0) +
    {(1-\rho_0)(1-2\rho_0)-m^2 \over \sqrt{(1-\rho_0)^2-m^2}} \cos\theta\right],
    \nonumber
\end{eqnarray}
where $\rho_i$ ($i=+,0,-$) is the fractional population of component $|i\rangle$,
($\sum_i \rho_i=1$), $m=\rho_+-\rho_-$ is the magnetization in a spin-1 Bose condensate,
a conserved quantity.
$\theta$ is the relative phase~\cite{Zhang05a}.
$\phi(\vec r)$ is a unit normalized spatial mode function
under the SMA determined from a
scalar Gross-Pitaevskii equation with an $s$-wave scattering length of $a_2$.
As before, the effective spin exchange interaction is given by
$c(t) = c_2(t)N\int d\vec r|\phi(\vec r)|^4$, albeit the time dependence,
with $N$ the total number of trapped atoms.
Although the system dynamics (\ref{eq:sma}) does not conserve the total spin energy
\begin{eqnarray}
{\cal E}(t) &=&
c(t)\rho_0\left[(1-\rho_0)+\sqrt{(1-\rho_0)^2-m^2}\cos\theta\right],
\label{eq:eng}
\end{eqnarray}
due to the temporal modulation, the transversal spin squared
$f_\perp^2 = f_x^2 + f_y^2 = 2{\cal E}(t)/c(t)$ remains conserved
and is determined solely by the initial condition.
Because ${\cal E}(t)$ and $c(t)$ are modulated exactly in the same manner,
replacing $\theta$ with $f_\perp^2$, equation (\ref{eq:sma})
is further simplified to
\begin{eqnarray}
(\dot \rho_0)^2 &=& {4c^2\over \hbar^2} f^2 (\rho_u-\rho_0)(\rho_0-\rho_d),
\end{eqnarray}
where
$
\rho_{u,d} = {f_\perp^2}\left(1 \pm
\sqrt{1-f^2}\right)/2f^2
$
with $f^2 = f_\perp^2+m^2$.
$\rho_{u(d)}$ takes the $+(-)$ sign, denoting the largest (smallest)
value of $\rho_0$ along the orbit and satisfies $\dot{\rho}_0|_{\rho_{u,d}} = 0$.
It is straightforward to find the solution
\begin{eqnarray}
\rho_0(t) &=& \frac{\rho_u+\rho_d}{2} +
\frac{\rho_u-\rho_d}{2}\sin[\gamma+ \int_0^t \beta(s) ds],
\label{eq:msma}
\end{eqnarray}
where $\beta(t) = \pm 2c(t)f/ \hbar$ is the frequency
of the periodic spin evolution and the $\pm$ sign denotes
the forward and backward evolutions, respectively, and
$$
\gamma = \mbox{atan}\left(\frac{\rho_0(0)-[(\rho_u+\rho_d)/2]}
{\sqrt{[\rho_u-\rho_0(0)][\rho_0(0)-\rho_d]}}\right)
$$
is given by the initial values of $\rho_0$ and $\theta$.
The above solution (\ref{eq:msma}) is valid for
an {\em arbitrary} temporal modulation function $c(t)$.

To control the spin dynamics, we consider several simple
but practical modulations in the following.
Based on these examples, we demonstrate that spin dynamics
with a modulated $c$ is very different from the free dynamics
and understand how a temporal modulated $c(t)$ affects the spin dynamics.

\begin{figure}
\includegraphics[width=3.25in]{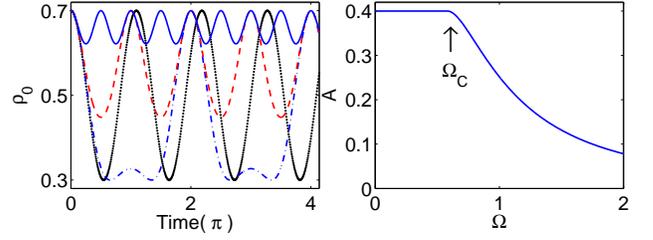}
\caption{(Color online) Left panel: Time dependent fractional population $\rho_0(t)$.
The black dotted line refers to the free evolution without modulation, while the blue dash-dotted,
red dashed, and blue solid lines are for $\Omega = 1/2$, $1$, and $2$, respectively.
The parameters and initial conditions are $\hbar=1$, $d=-1$, $m=0$, $\rho_0(0) = 0.7$, $\theta(0)=0$, and $\Omega_c \approx 0.58$.
Right panel: The dependence of oscillation amplitude $A$ of $\rho_0$ on the modulation frequency $\Omega$.}
\label{fig:cos} 
\end{figure}

First we consider a sinusoidal modulation with $c(t) = d \cos(\Omega t)$.
$d$ and $\Omega$ are respectively the modulation amplitude and frequency.
The solution Eq.~(\ref{eq:msma}) then becomes
\begin{eqnarray}
\rho_0(t) &=& \frac{\rho_u+\rho_d}{2} +
\frac{\rho_u-\rho_d}{2}\sin[\gamma + \eta \sin(\Omega t)],
\end{eqnarray}
with $ \eta = \pm 2df/\hbar\Omega $.
The corresponding results are illustrated in Fig.~\ref{fig:cos}
for $\Omega = 0$, $1/2$, $1$, and $2$.
The case of $\Omega=0$ is simply the free evolution without modulation
with a period $T_0=2\pi/\beta$ determined by the initial condition~\cite{Pu99}.
For other cases, irrespective of the values for $\Omega$,
the frequency of oscillation is always $\Omega$ and the
corresponding period is $2\pi/\Omega$.
As shown in Fig.~\ref{fig:cos}, the oscillation amplitudes show two distinctive regions:
one for $\Omega\le \Omega_c \equiv 2df / \pi\hbar$ (i.e. $\eta \ge \pi$) where the amplitude
is the same with/without modulation; another for $\Omega > \Omega_c$ (i.e. $\eta < \pi$) where the amplitude
decreases with modulation frequency $\Omega$. In this latter region,
it is easy to check that $A = (\rho_u-\rho_d)(1-\cos\eta)/2$ for the case shown in Fig.~\ref{fig:cos}.

\begin{figure}
\includegraphics[width=3.25in]{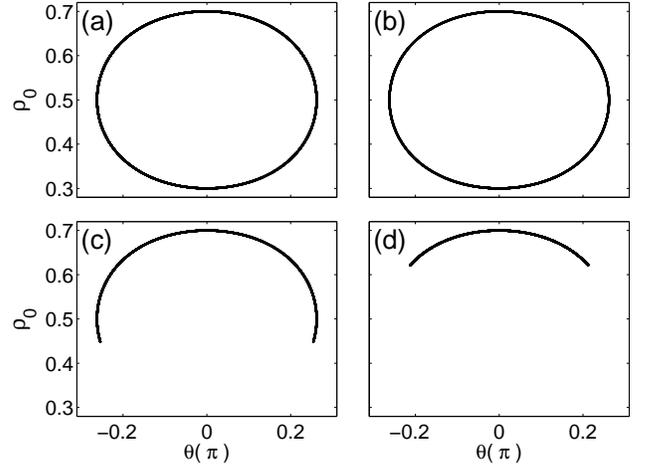}
\caption{ (a) Orbits without modulation;
Orbits with modulation for (b) $\Omega=1/2$, (c) $\Omega=1$, and (d) $\Omega=2$. Parameters are the same as in Fig.~\ref{fig:cos}.}
\label{fig:obt} 
\end{figure}

Figure~\ref{fig:obt} illustrates the orbits for the
corresponding spin dynamics. A full orbit is occupied if $\Omega < \Omega_c$
but only partial orbits are occupied when $\Omega > \Omega_c$.
The occupied portion deceases when $\Omega$ increases.
Spin dynamics for the first half period is reversed during the
second half period evolution, irrespective of the
values for $\Omega\neq 0$. This reversal is responsible for a more robust
modulated dynamics against various noises as noises are not reversed
and their effect can be averaged zero
according to coherent control theory~\cite{Slichter92}.

\begin{figure}
\includegraphics[width=3.25in]{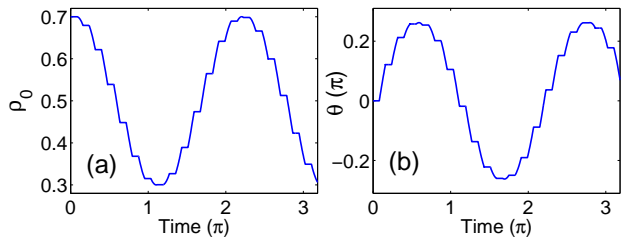}
\caption{(Color online) (a) Time-dependent fractional population $\rho_0$
and (b) time-dependent relative phase $\theta$ for the
periodic square modulation Eq.~(\ref{eq:sqr})
with $\tau=w=1/4$ and $d=-1$. }
\label{fig:sqr} 
\end{figure}

Next we consider a periodic square function modulation with
\begin{equation}
c(t) = \left\{\begin{array}{cc}d, & n(w+\tau)\le t < n(w+\tau)+w, \\ 0, &
n(w+\tau)+w \le t < (n+1)(w+\tau),
\end{array}\right. \label{eq:sqr}
\end{equation}
for $n=0,1,2,\cdots$. The spin dynamics is halted completely
if $n(w+\tau)+w \le t < (n+1)(w+\tau)$ and is unmodulated if $n(w+\tau)\le t < n(w+\tau)+w $.
The corresponding plots for $\rho_0$ and $\theta$ display
interesting step like features as shown in Figs.~\ref{fig:sqr}(a) and~\ref{fig:sqr}(b).

The modulation dynamics considered above offers
many interesting possibilities. A direct application
is to remove a dynamical instability observed in a ferromagnetically
interacting spin-1 condensate~\cite{Zhang05b, Chang05, Sadler06, Mur-Petit06, Gu07}.
This instability is removed whenever the imaginary part of the eigenfrequency
for the corresponding Bogoliubov excitation becomes zero in a modulation cycle.
We show this instability is indeed suppressed
in the following for the cosine modulation $c_2(t)=d\cos(\Omega t)$.
This suppression inhibits the spontaneous formation of spin domains.

We now consider our system of a homogeneous $^{87}$Rb spin-1 condensate
starting from an off-equilibrium initial state~\cite{Zhang05b}.
The averaged spin
$\vec f=\langle F_x\rangle \hat x + \langle F_y\rangle \hat y + \langle F_z\rangle \hat z$
is conserved where
$F_{x,y,z}$ are spin-1 matrices.
Starting from any stationary point,
the evolution of the collective excitations takes a compact form
\begin{eqnarray}
i\hbar \frac{\partial \vec x} {\partial t} = {\cal M}(t) \cdot \vec
x, \label{eq:bdg}
\end{eqnarray}
where $\vec x = (\delta \Psi_+, \delta \Psi_0, \delta \Psi_-, \delta \Psi_+^*, \delta \Psi_0^*, \delta \Psi_-^*)^T$
with deviations $\delta \Psi_j$ and $\delta \Psi_j^*$ from the stationary solution~\cite{Zhang05b}.

\begin{figure}
\includegraphics[width=3.25in]{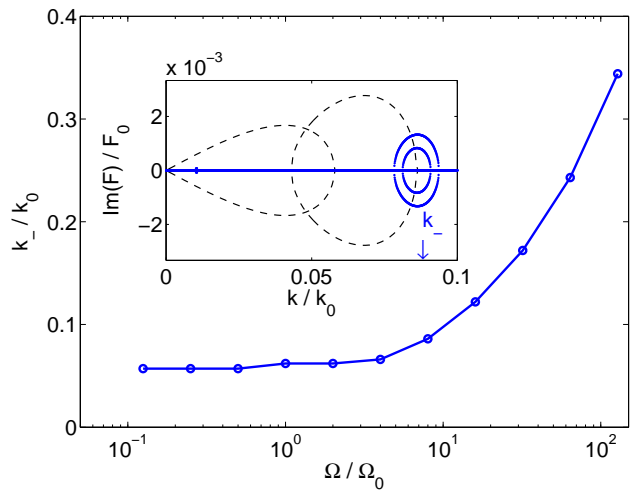}
\caption{(Color online) The dependence of the wave vector $k_-$ for the
most unstable mode on the modulation frequency $\Omega$. The inset illustrates
the imaginary part of a typical Bogoliubov spectrum under modulation
(blue double contours). For comparison, black dashed lines denote
the results without modulation.}
\label{fig:smi} 
\end{figure}

The general solution to Eq.~(\ref{eq:bdg}) is $\vec x(t) = U(t,0)\vec x(0)$
where $U(t_2,t_1) = {\cal T} \exp[-(i/\hbar) \int_{t_1}^{t_2} ds {\cal M}(s)]$
with ${\cal T}$ the time ordering operator.
For periodic modulation,
the solution during $t\in [pT,(p+1)T]$ is further simplified to
$\vec x(t) = U(t,pT) (U_T)^p \vec x(0)$, where $p=0,1,2,\cdots$
indexes the number of periods. $T=2\pi / \Omega$ is the period of modulation,
and $U_T=U(T,0)={\cal T} \exp[-(i/\hbar) \int_0^T ds {\cal M}(s)]$ is the evolution operator for a complete period.
$|\vec x(t)|$ will grow (decay) exponentially with $p$ if the modulus of $U_T$ are larger (smaller) than unity.
Diagonalizing $U_T$ and rewriting it as $U_T = V^\dag \exp(-iTF)V$,
where the Fluoquet operator $F$ is a 6-by-6 diagonal matrix,
the criteria for stable dynamics reduces to a vanishing imaginary part
of $F_q$ ($q=1,2,\cdots,6$). Unstable dynamics arises if the imaginary part is not zero,
while stable dynamics emerges if the imaginary part of all $F_q$ is exactly zero.

The inset of Fig.~\ref{fig:smi} shows the imaginary part of a typical spectrum for
the system under a cosine modulation of $c_2$.
We focus on the most unstable mode which in principle dominates the unstable dynamics.
Compared to the modulation free results (in dashed lines), we find the most unstable mode
(in double contours) is not only suppressed in amplitude but also shifted to larger wave vector.
The dependence of $k_-$ on the modulation frequency $\Omega$ is illustrated
in Fig.~\ref{fig:smi}. The almost independence of $k_-$ on
$\Omega$ at small values contrasts with a strong monotonic increase at large values of $\Omega$.

The emergence of spin domains is prohibited due to the modulation.
On one hand, the suppression of $F$ implies a smaller effective spin exchange interaction
thus a longer effective spin healing length $\xi$;
On the other hand, the up-shift of $k_-$ means shorter wave length $\lambda = 2\pi / k_-$.
If the unstable mode
wave length (potentially domain width) is smaller than the spin healing
length, the condensate is able to heal by itself.
The domain structure would never appear if $\xi$ exceeds $\lambda$.

Because of the modulation, however, the resulting dynamics becomes
completely different from the case of $c_2=0$ or no spin dynamics at all.
The modulation does not stop spin mixing dynamics, {\it i.e.},
as we continue to observe spin waves which is nominally disguised
in experiments by the spontaneously formed spin domains~\cite{Sadler06, Chang05}.
Furthermore, we expect the modulated spin dynamics to be robust against various
experimental noises because the periodic modulation effectively
cancels uncorrelated noises from alternating modulation periods.

\begin{figure}
\includegraphics[width=3.5in]{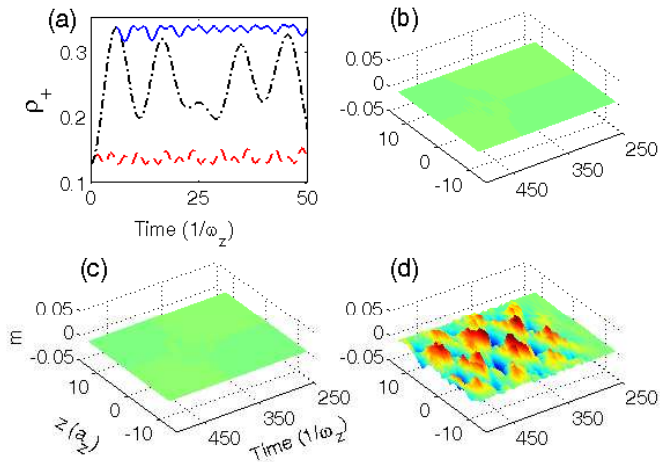}
\caption{(Color online) (a) Localization of the spin dynamics for a trapped
spin-1 condensate in 3 cases: (i) modulation starts at $t=0$ (red dashed line);
(ii) no modulation (black dash-dotted line); and (iii) modulation starts at $t=6$ (blue solid line). Spatial distribution $m(z)=
\int dr 2 \pi r (|\Phi_+(r,z)|^2-|\Phi_-(r,z)|^2)$ at different times for the above three cases: (b) --- case (i); (c) --- case (iii); (d) --- case (ii). $a_z=\sqrt{\hbar / M\omega_z}$. Trivial and flat $m(z)$ at early times ($t<250$) has been omitted. Dynamical instability induced spontaneous domain formation is prohibited by the modulated spin exchange interaction.}
\label{fig:lbec} 
\end{figure}

Finally we confirm the above conclusions for a trapped
spin-1 condensate with full numerical simulations.
We adopt experimental parameters as in
Ref.~\cite{Zhang05b}: The initial conditions are as in the experiment
\cite{Chang05}, with $^{87}$Rb condensates [$\rho_0(0)=0.744$, $\theta(0)=0$,
$N=2.0\times 10^5$, and $m=0$], in a trap $V_{\rm ext}(\vec r) =
(M/2)(\omega_x^2x^2 +\omega_y^2y^2+\omega_z^2z^2)$ with $\omega_x=\omega_y=
(2\pi) 240$ Hz and $\omega_z=(2\pi) 24$ Hz. The modulation function is $c_2(t)
= d \cos(\Omega t)$ with $d=c_2$, and $\Omega = \omega_z$ which is
about 3 times larger than the free spin evolution frequency $2\pi/T_0$.
Two cases will be considered: (1) The modulation is applied
immediately ($\rho_0$ oscillates around $\rho_u$); (2) The
modulation is turned on at $t=6\;(1/\omega_z)$ ($\rho_0$ oscillates around
$\rho_d$).

The results from numerical simulations are shown in Fig.~\ref{fig:lbec}.
With modulation, the spin dynamics is clearly localized as
$\rho_+$ (same for $\rho_-$) oscillates with a smaller amplitude around its initial value,
in contrast to the large amplitude unmodulated result [panel (a)].
In addition, the
unmodulated dynamics shows domain structures after $t\approx 300$,
due to the intrinsic dynamical instability. While for both
modulated cases, no spin domains are observed. Thus
temporal modulation of spin exchange interaction does suppress the intrinsic dynamical
instability and prohibit spontaneous domain formation.
In our extensive numerical simulations, we find that
when additional white noises are added, spin domains are found
to arise quicker for the unmodulated case; yet for
the two cases with modulations, almost the same behaviors are
observed as if no noise were added.

Although the life time of the condensate at optical Feshbach resonance is reduced dramatically in $^{87}$Rb gases~\cite{Hamley09}, we notice that there exists at least one magic window of relative frequency, e.g., between resonance $\beta$ and $\gamma$ in Fig.~7 of Ref.~\cite{Hamley09}, where the condensate life time lasts several hundred milliseconds and $c_2$ changes sign. On the other hand, the spin domain emerges in a time scale typically shorter than 100 ms~\cite{Leslie09}. Therefore the suppression of the domain formation in $^{87}$Rb condensate is experimentally feasible.

In summary, we propose to localize the spin mixing dynamics
in a spin-1 condensate by temporally modulating the spin exchange interaction.
For condensed atoms, the modulation can be facilitated with
the technique of optical Feshbach resonance \cite{Fatemi00,Hamley09}.
We demonstrate the suppression
of the intrinsic instability thus the inhibition of spontaneous spin domain
formation in a ferromagnetically interacting spin-1 Bose condensate,
such as $^{87}$Rb condensate in the $F=1$ manifold.
In addition, the effective freezing of spin mixing dynamics
due to spin exchange interaction provides a
cleaner approach to investigate
magnetic spin dipolar interaction effect in a $^{87}$Rb Bose condensate~\cite{Yi06, Vengalattore08}.

W. Z. acknowledges support by the 973 Program Grant No. 2009CB929300, the National Natural Science Foundation of China Grant No. 10904017, and the Program for New Century Excellent Talents in University.


\begin{thebibliography}{33}
\expandafter\ifx\csname natexlab\endcsname\relax\def\natexlab#1{#1}\fi
\expandafter\ifx\csname bibnamefont\endcsname\relax
  \def\bibnamefont#1{#1}\fi
\expandafter\ifx\csname bibfnamefont\endcsname\relax
  \def\bibfnamefont#1{#1}\fi
\expandafter\ifx\csname citenamefont\endcsname\relax
  \def\citenamefont#1{#1}\fi
\expandafter\ifx\csname url\endcsname\relax
  \def\url#1{\texttt{#1}}\fi
\expandafter\ifx\csname urlprefix\endcsname\relax\def\urlprefix{URL }\fi
\providecommand{\bibinfo}[2]{#2}
\providecommand{\eprint}[2][]{\url{#2}}

\bibitem[{\citenamefont{Landau and Lifshitz}(1960)}]{LandauBook60}
\bibinfo{author}{\bibfnamefont{L.~D.} \bibnamefont{Landau}} \bibnamefont{and}
  \bibinfo{author}{\bibfnamefont{E.~M.} \bibnamefont{Lifshitz}},
  \emph{\bibinfo{title}{Mechanics}} (\bibinfo{publisher}{Pergamon, Oxford},
  \bibinfo{year}{1960}).

\bibitem[{\citenamefont{Horvath et~al.}(1997)\citenamefont{Horvath, Thompson,
  and Knight}}]{Horvath97}
\bibinfo{author}{\bibfnamefont{G.}~\bibnamefont{Horvath}},
  \bibinfo{author}{\bibfnamefont{R.}~\bibnamefont{Thompson}}, \bibnamefont{and}
  \bibinfo{author}{\bibfnamefont{P.}~\bibnamefont{Knight}},
  \bibinfo{journal}{Contemporary Physics} \textbf{\bibinfo{volume}{38}},
  \bibinfo{pages}{25} (\bibinfo{year}{1997}).

\bibitem[{\citenamefont{Saito and Ueda}(2003)}]{Saito03}
\bibinfo{author}{\bibfnamefont{H.}~\bibnamefont{Saito}} \bibnamefont{and}
  \bibinfo{author}{\bibfnamefont{M.}~\bibnamefont{Ueda}},
  \bibinfo{journal}{Phys. Rev. Lett.} \textbf{\bibinfo{volume}{90}},
  \bibinfo{pages}{040403} (\bibinfo{year}{2003}).

\bibitem[{\citenamefont{Abdullaev and Kraenkel}(2000)}]{Abdullaev00}
\bibinfo{author}{\bibfnamefont{F.}~\bibnamefont{Abdullaev}} \bibnamefont{and}
  \bibinfo{author}{\bibfnamefont{R.}~\bibnamefont{Kraenkel}},
  \bibinfo{journal}{Phys. Lett. A} \textbf{\bibinfo{volume}{272}},
  \bibinfo{pages}{395 } (\bibinfo{year}{2000}).

\bibitem[{\citenamefont{Abdullaev et~al.}(2003)\citenamefont{Abdullaev, Caputo,
  Kraenkel, and Malomed}}]{Abdullaev03}
\bibinfo{author}{\bibfnamefont{F.~K.} \bibnamefont{Abdullaev}},
  \bibinfo{author}{\bibfnamefont{J.~G.} \bibnamefont{Caputo}},
  \bibinfo{author}{\bibfnamefont{R.~A.} \bibnamefont{Kraenkel}},
  \bibnamefont{and} \bibinfo{author}{\bibfnamefont{B.~A.}
  \bibnamefont{Malomed}}, \bibinfo{journal}{Phys. Rev. A}
  \textbf{\bibinfo{volume}{67}}, \bibinfo{pages}{013605}
  (\bibinfo{year}{2003}).

\bibitem[{\citenamefont{Montesinos et~al.}(2004)\citenamefont{Montesinos,
  P\'erez-Garc\'ia, and Michinel}}]{Montesinos04}
\bibinfo{author}{\bibfnamefont{G.~D.} \bibnamefont{Montesinos}},
  \bibinfo{author}{\bibfnamefont{V.~M.} \bibnamefont{P\'erez-Garc\'ia}},
  \bibnamefont{and} \bibinfo{author}{\bibfnamefont{H.}~\bibnamefont{Michinel}},
  \bibinfo{journal}{Phys. Rev. Lett.} \textbf{\bibinfo{volume}{92}},
  \bibinfo{pages}{133901} (\bibinfo{year}{2004}).

\bibitem[{\citenamefont{Zhang et~al.}(2005{\natexlab{a}})\citenamefont{Zhang,
  Zhou, Chang, Chapman, and You}}]{Zhang05b}
\bibinfo{author}{\bibfnamefont{W.}~\bibnamefont{Zhang}},
  \bibinfo{author}{\bibfnamefont{D.~L.} \bibnamefont{Zhou}},
  \bibinfo{author}{\bibfnamefont{M.-S.} \bibnamefont{Chang}},
  \bibinfo{author}{\bibfnamefont{M.~S.} \bibnamefont{Chapman}},
  \bibnamefont{and} \bibinfo{author}{\bibfnamefont{L.}~\bibnamefont{You}},
  \bibinfo{journal}{Phys. Rev. Lett.} \textbf{\bibinfo{volume}{95}},
  \bibinfo{eid}{180403} (\bibinfo{year}{2005}{\natexlab{a}}).

\bibitem[{\citenamefont{Chang et~al.}(2005)\citenamefont{Chang, Qin, Zhang,
  You, and Chapman}}]{Chang05}
\bibinfo{author}{\bibfnamefont{M.}~\bibnamefont{Chang}},
  \bibinfo{author}{\bibfnamefont{Q.}~\bibnamefont{Qin}},
  \bibinfo{author}{\bibfnamefont{W.}~\bibnamefont{Zhang}},
  \bibinfo{author}{\bibfnamefont{L.}~\bibnamefont{You}}, \bibnamefont{and}
  \bibinfo{author}{\bibfnamefont{M.}~\bibnamefont{Chapman}},
  \bibinfo{journal}{Nature Phys.} \textbf{\bibinfo{volume}{1}},
  \bibinfo{pages}{111} (\bibinfo{year}{2005}).

\bibitem[{\citenamefont{Sadler et~al.}(2006)\citenamefont{Sadler, Higbie,
  Leslie, Vengalattore, and Stamper-Kurn}}]{Sadler06}
\bibinfo{author}{\bibfnamefont{L.~E.} \bibnamefont{Sadler}},
  \bibinfo{author}{\bibfnamefont{J.~M.} \bibnamefont{Higbie}},
  \bibinfo{author}{\bibfnamefont{S.~R.} \bibnamefont{Leslie}},
  \bibinfo{author}{\bibfnamefont{M.}~\bibnamefont{Vengalattore}},
  \bibnamefont{and} \bibinfo{author}{\bibfnamefont{D.~M.}
  \bibnamefont{Stamper-Kurn}}, \bibinfo{journal}{Nature (London)}
  \textbf{\bibinfo{volume}{443}}, \bibinfo{pages}{312} (\bibinfo{year}{2006}).

\bibitem[{\citenamefont{Vengalattore et~al.}(2007)\citenamefont{Vengalattore,
  Higbie, Leslie, Guzman, Sadler, and Stamper-Kurn}}]{Vengalattore07}
\bibinfo{author}{\bibfnamefont{M.}~\bibnamefont{Vengalattore}},
  \bibinfo{author}{\bibfnamefont{J.~M.} \bibnamefont{Higbie}},
  \bibinfo{author}{\bibfnamefont{S.~R.} \bibnamefont{Leslie}},
  \bibinfo{author}{\bibfnamefont{J.}~\bibnamefont{Guzman}},
  \bibinfo{author}{\bibfnamefont{L.~E.} \bibnamefont{Sadler}},
  \bibnamefont{and} \bibinfo{author}{\bibfnamefont{D.~M.}
  \bibnamefont{Stamper-Kurn}}, \bibinfo{journal}{Phys. Rev. Lett.}
  \textbf{\bibinfo{volume}{98}}, \bibinfo{eid}{200801} (\bibinfo{year}{2007}).

\bibitem[{\citenamefont{Yi and Pu}(2006)}]{Yi06}
\bibinfo{author}{\bibfnamefont{S.}~\bibnamefont{Yi}} \bibnamefont{and}
  \bibinfo{author}{\bibfnamefont{H.}~\bibnamefont{Pu}}, \bibinfo{journal}{Phys.
  Rev. Lett.} \textbf{\bibinfo{volume}{97}}, \bibinfo{eid}{020401}
  (\bibinfo{year}{2006}).

\bibitem[{\citenamefont{Vengalattore et~al.}(2008)\citenamefont{Vengalattore,
  Leslie, Guzman, and Stamper-Kurn}}]{Vengalattore08}
\bibinfo{author}{\bibfnamefont{M.}~\bibnamefont{Vengalattore}},
  \bibinfo{author}{\bibfnamefont{S.~R.} \bibnamefont{Leslie}},
  \bibinfo{author}{\bibfnamefont{J.}~\bibnamefont{Guzman}}, \bibnamefont{and}
  \bibinfo{author}{\bibfnamefont{D.~M.} \bibnamefont{Stamper-Kurn}},
  \bibinfo{journal}{Phys. Rev. Lett.} \textbf{\bibinfo{volume}{100}},
  \bibinfo{eid}{170403} (\bibinfo{year}{2008}).

\bibitem[{\citenamefont{Fatemi et~al.}(2000)\citenamefont{Fatemi, Jones, and
  Lett}}]{Fatemi00}
\bibinfo{author}{\bibfnamefont{F.~K.} \bibnamefont{Fatemi}},
  \bibinfo{author}{\bibfnamefont{K.~M.} \bibnamefont{Jones}}, \bibnamefont{and}
  \bibinfo{author}{\bibfnamefont{P.~D.} \bibnamefont{Lett}},
  \bibinfo{journal}{Phys. Rev. Lett.} \textbf{\bibinfo{volume}{85}},
  \bibinfo{pages}{4462} (\bibinfo{year}{2000}).

\bibitem[{\citenamefont{Cheng et~al.}(2008)\citenamefont{Cheng, Jing, and
  Yan}}]{Cheng08}
\bibinfo{author}{\bibfnamefont{J.}~\bibnamefont{Cheng}},
  \bibinfo{author}{\bibfnamefont{H.}~\bibnamefont{Jing}}, \bibnamefont{and}
  \bibinfo{author}{\bibfnamefont{Y.}~\bibnamefont{Yan}},
  \bibinfo{journal}{Phys. Rev. A} \textbf{\bibinfo{volume}{77}},
  \bibinfo{pages}{061604} (\bibinfo{year}{2008}).

\bibitem[{\citenamefont{Hamley et~al.}(2009)\citenamefont{Hamley, Bookjans,
  Behin-Aein, Ahmadi, and Chapman}}]{Hamley09}
\bibinfo{author}{\bibfnamefont{C.~D.} \bibnamefont{Hamley}},
  \bibinfo{author}{\bibfnamefont{E.~M.} \bibnamefont{Bookjans}},
  \bibinfo{author}{\bibfnamefont{G.}~\bibnamefont{Behin-Aein}},
  \bibinfo{author}{\bibfnamefont{P.}~\bibnamefont{Ahmadi}}, \bibnamefont{and}
  \bibinfo{author}{\bibfnamefont{M.~S.} \bibnamefont{Chapman}},
  \bibinfo{journal}{Phys. Rev. A} \textbf{\bibinfo{volume}{79}},
  \bibinfo{eid}{023401} (\bibinfo{year}{2009}).

\bibitem[{\citenamefont{Jing et~al.}(2009)\citenamefont{Jing, Fu, Geng, and
  Liu}}]{Jing09}
\bibinfo{author}{\bibfnamefont{H.}~\bibnamefont{Jing}},
  \bibinfo{author}{\bibfnamefont{J.}~\bibnamefont{Fu}},
  \bibinfo{author}{\bibfnamefont{Z.}~\bibnamefont{Geng}}, \bibnamefont{and}
  \bibinfo{author}{\bibfnamefont{W.-M.} \bibnamefont{Liu}},
  \bibinfo{journal}{Phys. Rev. A} \textbf{\bibinfo{volume}{79}},
  \bibinfo{eid}{045601} (\bibinfo{year}{2009}).

\bibitem[{\citenamefont{Jack and Yamashita}(2005)}]{Jack05}
\bibinfo{author}{\bibfnamefont{M.~W.} \bibnamefont{Jack}} \bibnamefont{and}
  \bibinfo{author}{\bibfnamefont{M.}~\bibnamefont{Yamashita}},
  \bibinfo{journal}{Phys. Rev. A} \textbf{\bibinfo{volume}{71}},
  \bibinfo{pages}{033619} (\bibinfo{year}{2005}).

\bibitem[{\citenamefont{Theis et~al.}(2004)\citenamefont{Theis, Thalhammer,
  Winkler, Hellwig, Ruff, Grimm, and Denschlag}}]{Theis04}
\bibinfo{author}{\bibfnamefont{M.}~\bibnamefont{Theis}},
  \bibinfo{author}{\bibfnamefont{G.}~\bibnamefont{Thalhammer}},
  \bibinfo{author}{\bibfnamefont{K.}~\bibnamefont{Winkler}},
  \bibinfo{author}{\bibfnamefont{M.}~\bibnamefont{Hellwig}},
  \bibinfo{author}{\bibfnamefont{G.}~\bibnamefont{Ruff}},
  \bibinfo{author}{\bibfnamefont{R.}~\bibnamefont{Grimm}}, \bibnamefont{and}
  \bibinfo{author}{\bibfnamefont{J.~H.} \bibnamefont{Denschlag}},
  \bibinfo{journal}{Phys. Rev. Lett.} \textbf{\bibinfo{volume}{93}},
  \bibinfo{pages}{123001} (\bibinfo{year}{2004}).

\bibitem[{\citenamefont{Theis}()}]{Theis05}
\bibinfo{author}{\bibfnamefont{M.}~\bibnamefont{Theis}},
  \emph{\bibinfo{title}{Optical feshbach resonances in a bose-einstein
  condensate}}, \bibinfo{note}{ph. D. thesis, Universit\:at Innsbruck, Austria
  (2005)}.

\bibitem[{\citenamefont{Ho}(1998)}]{Ho98}
\bibinfo{author}{\bibfnamefont{T.-L.} \bibnamefont{Ho}},
  \bibinfo{journal}{Phys. Rev. Lett.} \textbf{\bibinfo{volume}{81}},
  \bibinfo{pages}{742} (\bibinfo{year}{1998}).

\bibitem[{\citenamefont{Law et~al.}(1998)\citenamefont{Law, Pu, and
  Bigelow}}]{Law98}
\bibinfo{author}{\bibfnamefont{C.~K.} \bibnamefont{Law}},
  \bibinfo{author}{\bibfnamefont{H.}~\bibnamefont{Pu}}, \bibnamefont{and}
  \bibinfo{author}{\bibfnamefont{N.~P.} \bibnamefont{Bigelow}},
  \bibinfo{journal}{Phys. Rev. Lett.} \textbf{\bibinfo{volume}{81}},
  \bibinfo{pages}{5257} (\bibinfo{year}{1998}).

\bibitem[{\citenamefont{Ohmi and Machida}(1998)}]{Ohmi98}
\bibinfo{author}{\bibfnamefont{T.}~\bibnamefont{Ohmi}} \bibnamefont{and}
  \bibinfo{author}{\bibfnamefont{K.}~\bibnamefont{Machida}},
  \bibinfo{journal}{J. Phys. Soc. Jpn} \textbf{\bibinfo{volume}{67}},
  \bibinfo{pages}{1822} (\bibinfo{year}{1998}).

\bibitem[{\citenamefont{Yi et~al.}(2002)\citenamefont{Yi, M\"ustecapl\ifmmode
  \imath \else \i \fi{}o\ifmmode~\breve{g}\else \u{g}\fi{}lu, Sun, and
  You}}]{Yi02}
\bibinfo{author}{\bibfnamefont{S.}~\bibnamefont{Yi}},
  \bibinfo{author}{\bibfnamefont{O.~E.} \bibnamefont{M\"ustecapl\ifmmode \imath
  \else \i \fi{}o\ifmmode~\breve{g}\else \u{g}\fi{}lu}},
  \bibinfo{author}{\bibfnamefont{C.~P.} \bibnamefont{Sun}}, \bibnamefont{and}
  \bibinfo{author}{\bibfnamefont{L.}~\bibnamefont{You}},
  \bibinfo{journal}{Phys. Rev. A} \textbf{\bibinfo{volume}{66}},
  \bibinfo{pages}{011601(R)} (\bibinfo{year}{2002}).

\bibitem[{\citenamefont{Zhang et~al.}(2003)\citenamefont{Zhang, Yi, and
  You}}]{Zhang03}
\bibinfo{author}{\bibfnamefont{W.}~\bibnamefont{Zhang}},
  \bibinfo{author}{\bibfnamefont{S.}~\bibnamefont{Yi}}, \bibnamefont{and}
  \bibinfo{author}{\bibfnamefont{L.}~\bibnamefont{You}}, \bibinfo{journal}{New
  J. Phys.} \textbf{\bibinfo{volume}{5}}, \bibinfo{pages}{77}
  (\bibinfo{year}{2003}).

\bibitem[{\citenamefont{Black et~al.}(2007)\citenamefont{Black, Gomez, Turner,
  Jung, and Lett}}]{Black07}
\bibinfo{author}{\bibfnamefont{A.~T.} \bibnamefont{Black}},
  \bibinfo{author}{\bibfnamefont{E.}~\bibnamefont{Gomez}},
  \bibinfo{author}{\bibfnamefont{L.~D.} \bibnamefont{Turner}},
  \bibinfo{author}{\bibfnamefont{S.}~\bibnamefont{Jung}}, \bibnamefont{and}
  \bibinfo{author}{\bibfnamefont{P.~D.} \bibnamefont{Lett}},
  \bibinfo{journal}{Phys. Rev. Lett.} \textbf{\bibinfo{volume}{99}},
  \bibinfo{eid}{070403} (\bibinfo{year}{2007}).

\bibitem[{\citenamefont{Liu et~al.}(2009{\natexlab{a}})\citenamefont{Liu, Jung,
  Maxwell, Turner, Tiesinga, and Lett}}]{Liu09a}
\bibinfo{author}{\bibfnamefont{Y.}~\bibnamefont{Liu}},
  \bibinfo{author}{\bibfnamefont{S.}~\bibnamefont{Jung}},
  \bibinfo{author}{\bibfnamefont{S.~E.} \bibnamefont{Maxwell}},
  \bibinfo{author}{\bibfnamefont{L.~D.} \bibnamefont{Turner}},
  \bibinfo{author}{\bibfnamefont{E.}~\bibnamefont{Tiesinga}}, \bibnamefont{and}
  \bibinfo{author}{\bibfnamefont{P.~D.} \bibnamefont{Lett}},
  \bibinfo{journal}{Phys. Rev. Lett.} \textbf{\bibinfo{volume}{102}},
  \bibinfo{eid}{125301} (\bibinfo{year}{2009}{\natexlab{a}}).

\bibitem[{\citenamefont{Liu et~al.}(2009{\natexlab{b}})\citenamefont{Liu,
  Gomez, Maxwell, Turner, Tiesinga, and Lett}}]{Liu09b}
\bibinfo{author}{\bibfnamefont{Y.}~\bibnamefont{Liu}},
  \bibinfo{author}{\bibfnamefont{E.}~\bibnamefont{Gomez}},
  \bibinfo{author}{\bibfnamefont{S.~E.} \bibnamefont{Maxwell}},
  \bibinfo{author}{\bibfnamefont{L.~D.} \bibnamefont{Turner}},
  \bibinfo{author}{\bibfnamefont{E.}~\bibnamefont{Tiesinga}}, \bibnamefont{and}
  \bibinfo{author}{\bibfnamefont{P.~D.} \bibnamefont{Lett}},
  \bibinfo{journal}{Phys. Rev. Lett.} \textbf{\bibinfo{volume}{102}},
  \bibinfo{eid}{225301} (\bibinfo{year}{2009}{\natexlab{b}}).

\bibitem[{\citenamefont{Zhang et~al.}(2005{\natexlab{b}})\citenamefont{Zhang,
  Zhou, Chang, Chapman, and You}}]{Zhang05a}
\bibinfo{author}{\bibfnamefont{W.}~\bibnamefont{Zhang}},
  \bibinfo{author}{\bibfnamefont{D.~L.} \bibnamefont{Zhou}},
  \bibinfo{author}{\bibfnamefont{M.-S.} \bibnamefont{Chang}},
  \bibinfo{author}{\bibfnamefont{M.~S.} \bibnamefont{Chapman}},
  \bibnamefont{and} \bibinfo{author}{\bibfnamefont{L.}~\bibnamefont{You}},
  \bibinfo{journal}{Phys. Rev. A} \textbf{\bibinfo{volume}{72}},
  \bibinfo{eid}{013602} (\bibinfo{year}{2005}{\natexlab{b}}).

\bibitem[{\citenamefont{Pu et~al.}(1999)\citenamefont{Pu, Law, Raghavan,
  Eberly, and Bigelow}}]{Pu99}
\bibinfo{author}{\bibfnamefont{H.}~\bibnamefont{Pu}},
  \bibinfo{author}{\bibfnamefont{C.~K.} \bibnamefont{Law}},
  \bibinfo{author}{\bibfnamefont{S.}~\bibnamefont{Raghavan}},
  \bibinfo{author}{\bibfnamefont{J.~H.} \bibnamefont{Eberly}},
  \bibnamefont{and} \bibinfo{author}{\bibfnamefont{N.~P.}
  \bibnamefont{Bigelow}}, \bibinfo{journal}{Phys. Rev. A}
  \textbf{\bibinfo{volume}{60}}, \bibinfo{pages}{1463} (\bibinfo{year}{1999}).

\bibitem[{\citenamefont{Slichter}(1992)}]{Slichter92}
\bibinfo{author}{\bibfnamefont{C.~P.} \bibnamefont{Slichter}},
  \emph{\bibinfo{title}{Principles of Magnetic Resonance}}
  (\bibinfo{publisher}{Springer-Verlag, New York}, \bibinfo{year}{1992}).

\bibitem[{\citenamefont{Mur-Petit et~al.}(2006)\citenamefont{Mur-Petit,
  Guilleumas, Polls, Sanpera, Lewenstein, Bongs, and Sengstock}}]{Mur-Petit06}
\bibinfo{author}{\bibfnamefont{J.}~\bibnamefont{Mur-Petit}},
  \bibinfo{author}{\bibfnamefont{M.}~\bibnamefont{Guilleumas}},
  \bibinfo{author}{\bibfnamefont{A.}~\bibnamefont{Polls}},
  \bibinfo{author}{\bibfnamefont{A.}~\bibnamefont{Sanpera}},
  \bibinfo{author}{\bibfnamefont{M.}~\bibnamefont{Lewenstein}},
  \bibinfo{author}{\bibfnamefont{K.}~\bibnamefont{Bongs}}, \bibnamefont{and}
  \bibinfo{author}{\bibfnamefont{K.}~\bibnamefont{Sengstock}},
  \bibinfo{journal}{Phys. Rev. A} \textbf{\bibinfo{volume}{73}},
  \bibinfo{eid}{013629} (\bibinfo{year}{2006}).

\bibitem[{\citenamefont{Gu and Qiu}(2007)}]{Gu07}
\bibinfo{author}{\bibfnamefont{Q.}~\bibnamefont{Gu}} \bibnamefont{and}
  \bibinfo{author}{\bibfnamefont{H.}~\bibnamefont{Qiu}},
  \bibinfo{journal}{Phys. Rev. Lett.} \textbf{\bibinfo{volume}{98}},
  \bibinfo{eid}{200401} (\bibinfo{year}{2007}).

\bibitem[{\citenamefont{Leslie et~al.}(2009)\citenamefont{Leslie, Guzman,
  Vengalattore, Sau, Cohen, and Stamper-Kurn}}]{Leslie09}
\bibinfo{author}{\bibfnamefont{S.~R.} \bibnamefont{Leslie}},
  \bibinfo{author}{\bibfnamefont{J.}~\bibnamefont{Guzman}},
  \bibinfo{author}{\bibfnamefont{M.}~\bibnamefont{Vengalattore}},
  \bibinfo{author}{\bibfnamefont{J.~D.} \bibnamefont{Sau}},
  \bibinfo{author}{\bibfnamefont{M.~L.} \bibnamefont{Cohen}}, \bibnamefont{and}
  \bibinfo{author}{\bibfnamefont{D.~M.} \bibnamefont{Stamper-Kurn}},
  \bibinfo{journal}{Phys. Rev. A} \textbf{\bibinfo{volume}{79}},
  \bibinfo{pages}{043631} (\bibinfo{year}{2009}).

\end{thebibliography}

\end{document}